\documentclass[a4paper]{jpconf}
\usepackage{graphicx}
\usepackage{epsfig}
\usepackage[latin1]{inputenc}
\def\be{\begin{equation}}
\def\ee{\end{equation}}
\def\ba{\begin{array}}
\def\bacc{\begin{array} {cc}}
\def\ea{\end{array}}
\def\bea{\begin{eqnarray}}
\def\eea{\end{eqnarray}}
\def\bd{\begin{displaymath}}
\def\ed{\end{displaymath}}

\begin{document}

\hspace{12cm} UAB-FT-667

\title{CPT, Lorentz invariance and anomalous clash of symmetries}

\author{Alberto Salvio}

\address
{Institut de Th\'eorie des Ph\'enom\`enes Physiques, EPFL, CH-1015
Lausanne, Switzerland  \\ \vspace{0.05cm} 
and  IFAE, Universitat Aut\`{o}noma de
Barcelona, 08193 Bellaterra, Barcelona, Spain}

\ead{salvio@ifae.es}

\begin{abstract}
In this paper we first discuss the analysis regarding the role of
Lorentz symmetry in the perturbative non-gravitational anomalies for
a family of fermions, which has been recently performed in {\tt
arXiv:0809.0184}. The theory is assumed to be translational
invariant, power-counting renormalizable and based on a local
action, but is allowed to have general Lorentz violating operators,
including those that break CPT. The main result is that Lorentz
symmetry does not participate in the clash of symmetries that leads
to the anomalies. Moreover, here we provide a simple {\it semiclassical}
argument that shortly illustrates the origin of this fact.
\end{abstract}

\section{Introduction}
Field theories that do not possess Lorentz symmetry have attracted
much interest among particle physicists during the last decades. One
of the main motivations is the possibility to interpret
these models as effective descriptions of more fundamental theories
where gravity is consistently included and the breaking of Lorentz
invariance occurs spontaneously \cite{Wald:1980nm}. Relaxing Lorentz
invariance might also open further ways to address phenomenological
problems (some examples are provided in \cite{Bertolami:1996cq}).

Among the most popular frameworks, which extend the Standard Model
(SM) by incorporating Lorentz violating operators, there are the
works by Coleman and Glashow \cite{Coleman:1998ti} and Colladay and
Kostelecky \cite{Colladay:1996iz,Colladay:1998fq}, where quite
general assumptions, like locality, translational invariance and
power-counting renormalizability, are made. This Standard Model
Extension (SME) provides a framework where Lorentz symmetry can be
tested experimentally and bounds on the Lorentz violating operators
can be obtained explicitly. Also such a framework gives us the
possibility of investigating conceptual issues related to the role
of Lorentz invariance in modern theories of Particle Physics. Much
work has been done along these lines. For example causality,
stability \cite{Kostelecky:2000mm}, renormalization
\cite{Kostelecky:2001jc,Colladay:2006rk} and gauge invariance in the
extended QED \cite{Jackiw:1999yp}  have been studied in the presence
of small Lorentz violating perturbations.

In the present paper we first summarize the analysis of Ref.
\cite{Salvio:2008ta}. There a derivation of the perturbative
non-gravitational anomalies in the presence of quite general (but
small) Lorentz violating operators  has been provided, including the
violation of CPT. What is the role of Lorentz invariance in the
anomalies? Is it possible to relax the standard anomaly cancellation
conditions by relaxing Lorentz symmetry? These questions can be
reformulated as follows. Are the anomalies the impossibility of
defining a quantum {\it relativistic} theory with simultaneously conserved
internal currents or does Lorentz symmetry behave as a spectator in
this clash? The results of Ref. \cite{Salvio:2008ta} show that the
second possibility is the correct one, at least by making the
general assumptions of the SME and  by assuming the mixing between
fermions to be diagonal in the family space. Here we also provide a
further argument which leads to the same conclusion, but has the
advantage of being relatively simple and short. As we shall see,
this reasoning exploits the structure of the classical fermion
action in the presence of Lorentz violating parameters and the
symmetries of the anomalous Ward identities in standard theories.

\section{The fermion sector and general results}

In the following we will write an action that contains a certain
number of Lorentz violating terms. Let us assume the corresponding
parameters to be small (in a sense that will be clarified later on).
This assumption allows us to consider the fermion field as an object
with four {\it spinorial} components, as a discrete quantity cannot
be changed by a small perturbation. The first part of this section
is completely standard and has been introduced to fix the
conventions for the subsequent parts. We consider a family of
fermions $\psi$ in a general representation of a (compact Lie) group
$G$. We take as part of the definition of family the property that
the representation can be made by one or more than one irreducible
representations (irreps), but the irreps are all different.   The
infinitesimal action of $G$ on $\psi$ is
\be \delta \psi = i \Omega \psi\equiv i \Omega^b \left(t^b_L P_L  +
t^b_R P_R\right)\psi, \label{psitransf}\ee
where $\Omega^b$ represent the group transformation parameters,
$P_{L(R)}\equiv \left(1\pm \gamma_5\right)/2$ is the projector on
the left-handed (right-handed) subspaces, $\gamma_5= i\gamma_0
\gamma_1 \gamma_2 \gamma_3$ and $t^b_{L(R)}$ are the hermitian
generators in the left-handed (right-handed) representation.

We imagine that each generator of $G$ corresponds to a gauge field
$A_{\mu}^b$, $\mu=0,1,2,3$, but we do not require each gauge field
to be dynamical: in this way we can study both the anomalies
associated with gauge currents and those associated with global
currents. The action of $G$ on $A_{\mu}^b$ and the covariant
derivative of the fermions are respectively defined by
\be \delta A_{\mu}^b =  f^{c d b} \Omega^d A_{\mu}^c +
\partial_{\mu} \Omega^b, \label{Atransf}\ee
where $f^{c d b}$ represents the structure constants of $G$,
satisfying $[t^b_L,t^c_L]=if^{bcd}t^d_L$, and $$ D_{\mu}\psi \equiv
\left[\partial_{\mu} -iA_{\mu}^b\left(t^b_L P_L  + t^b_R P_R\right)
\right]\psi,$$ like in Lorentz invariant theories, as their form
comes from gauge invariance and therefore is insensitive to any
Lorentz violation. Here we only consider chiral representations:
\be t^b_L \neq t^b_R, \quad \mbox{for some $b$}.\label{chirality}\ee
This is indeed the only case when anomalies can appear in Lorentz
invariant theories.

All the ingredients introduced so far are also present in standard
theories. We now want to write a (classical) action for $\psi$ in
the $A^b_{\mu}$ background, which involves Lorentz violating terms.
By following the works of Coleman and Glashow \cite{Coleman:1998ti}
and Colladay and Kostelecky \cite{Colladay:1996iz,Colladay:1998fq},
we assume the following properties:
\begin{itemize}
\item Locality,
 \item Translational invariance,
\item Power-counting renormalizability (operators with dimension greater than four are not allowed).
\end{itemize}
These requirements tell us that the general form of the action is
\cite{Colladay:1998fq}
\be S= \int d^4x
\left(i\,\overline{\psi}\Gamma^{\mu} D_{\mu} \psi - \overline{\psi}
M\psi \right),\label{action}\ee
where we have adopted the signature $\eta_{\mu \nu}=$
diag$(+1,-1,-1,-1)$, $\Gamma^{\mu}$ and $M$ are general constant
$4\times 4$ matrices:
\bea \Gamma^{\mu}&\equiv &c^{\mu}_{\,\,\,\nu}\gamma^{\nu} +d^{\mu}_{\,\,\,\nu}\gamma_5 \gamma^{\nu} + e^{\mu} + i f^{\mu} \gamma_5 +\frac{1}{2} g^{\mu \nu \rho} \sigma_{\nu \rho}, \\
M&\equiv &m + \frac{1}{2}H^{\mu \nu} \sigma_{\mu \nu} +
a_{\mu}\gamma^{\mu} +b_{\mu} \gamma_5 \gamma^{\mu}. \label{Mdef}\eea
Observe however that an additional term of the form $m'\gamma_5$ can
be added to (\ref{Mdef}), but this may be removed from the action
via a chiral transformation. Here $\gamma^{\mu}$ are the usual Dirac
matrices, $\sigma_{\mu \nu}\equiv i[\gamma_{\mu},\gamma_{\nu}]/4$
and we have introduced the Lorentz violating parameters
\be c^{\mu}_{\,\,\,\nu}-\delta^{\mu}_{\nu}, d^{\mu}_{\,\,\,\nu},
e^{\mu}, f^{\mu}, g^{\mu \nu \rho}, H^{\mu \nu}, a_{\mu},
b_{\mu}.\label{LVparam}\ee
If the fermion representation is made by more than one irreps, the
parameters in (\ref{LVparam}) are generically different for
different irreps; we understand here an additional index labeling
different irreps. The Lorentz violating perturbations that we have
introduced can be divided into a {\it CPT-even} set $$c^{\mu}_{\,\,\,\nu}-
\delta^{\mu}_{\nu}, \,\,\, d^{\mu}_{\,\,\,\nu}, \,\,\, H^{\mu \nu}$$ and a
{\it CPT-odd} one $$e^{\mu},\,\,\,  f^{\mu},\,\,\,  g^{\mu \nu \rho}, \,\,\,a_{\mu},
\,\,\,b_{\mu}.$$ The experimental limits (for a recent summary of
experimental constraints see \cite{Kostelecky:2008ts}) require that,
in a frame in which the earth is not relativistic, all the
quantities in (\ref{LVparam}) are very small, in the sense that
$c^{\mu}_{\,\,\,\nu}-\delta^{\mu}_{\nu}, d^{\mu}_{\,\,\,\nu},
e^{\mu}, f^{\mu}, g^{\mu \nu \rho}<<1$ and  $H^{\mu \nu}, a_{\mu},
b_{\mu}<< m$ \cite{Kostelecky:2000mm}. In this paper we always work
in such a frame.   Also the parameters in (\ref{LVparam}) and $m$
are real as a consequence of $S^{\dagger}=S$ (in our conventions
$\overline{\psi}\equiv \psi^{\dagger}\gamma^0$ and
$\left(\gamma^{\mu}\right)^{\dagger}=\gamma^0 \gamma^{\mu}
\gamma^0$).

Some consistency checks of this model (in the free field case,
$A_{\mu}^b=0$) have been performed in \cite{Kostelecky:2000mm}. There
it is shown that inconsistencies emerge at very high energies or
equivalently in frames that move at very high speed with respect to
earth-based laboratories. These energies (or equivalently boosts)
are at a very high scale $\Lambda$ where the spontaneous symmetry
breaking of Lorentz invariance occurs. For example it is
conceivable, but not obligatory, that $\Lambda$ is the Planck scale.
Therefore, the model at hand should be considered as a low energy
effective description. From an effective field theory point of view
we expect \cite{Kostelecky:2000mm}
$c^{\mu}_{\,\,\,\nu}-\delta^{\mu}_{\nu}, d^{\mu}_{\,\,\,\nu},
e^{\mu}, f^{\mu}, g^{\mu \nu \rho}$ to be at most of order
$m/\Lambda$ and $H^{\mu \nu}, a_{\mu}, b_{\mu}$ to be at most of
order $m^2/\Lambda$ and therefore these parameters are tiny if $m$
is identified with the mass of the observed fermions.

In the following we will not assume $-\overline{\psi}M\psi$ to be
invariant under (\ref{psitransf}); in this way our analysis will be
applicable also to those theories, like the minimal SM, where the
fermion masses emerge from the spontaneous symmetry breaking of a
gauge symmetry. However, we do assume the first term in
(\ref{action}) to be invariant under (\ref{psitransf}) as, at least
in the power-counting renormalizable case, the Higgs mechanism
cannot modify that term. Since the generators satisfy
(\ref{chirality}), we have
\be \Gamma^{\mu}=c^{\mu}_{\,\,\,\nu}\gamma^{\nu}
+d^{\mu}_{\,\,\,\nu}\gamma_5 \gamma^{\nu}, \label{Gansatz}\ee
which is also the most general form of $\Gamma^{\mu}$ compatible
with the SM gauge group \cite{Colladay:1998fq}.

To study anomalies we introduce the functional $W[A]$ in the
standard way, that is $\exp\left(i\,W[A]\right) \equiv \int \delta
\psi \delta \overline{\psi} \,\exp\left( i\, S[A] \right)$, with the
normalization of the fermion measure chosen in a way that
$\exp\left(i\,W[0]\right)=1$. As usual the absence of anomalies
corresponds  to the gauge invariance of $W[A]$ under
(\ref{Atransf}):
\be \delta W[A] = 0 \quad + \,\,\,\mbox{$M$-terms}, \quad \mbox{(in
the absence of anomalies)},\label{no-anomal}\ee
where  $M$-terms represent the non invariance of $W[A]$ due to non
gauge invariant terms in $-\overline{\psi}M\psi$, if any. Condition
(\ref{no-anomal}) is equivalent to the Ward identities (WIs) for the
n-point Green functions
\be \langle J^{\mu_1}_{b_1}(x_1) ...
J^{\mu_n}_{b_n}(x_n)\rangle=\int \delta \psi \delta
\overline{\psi}\,\exp\left( i\, S[A=0] \right)J^{\mu_1}_{b_1}(x_1)
... J^{\mu_n}_{b_n}(x_n), \label{Green}\ee
like in the Lorentz invariant case. However, here we have to change
the definition of the currents according to our classical action:
$$J^{\mu}_b \equiv \overline{\psi}\,  \Gamma^{\mu} T_b \psi, \quad \mbox{with}\quad
T_b \equiv t^b_L P_L  + t^b_R P_R.$$

We now summarize the physical results of Ref. \cite{Salvio:2008ta}.
The WIs for (\ref{Green}) can be derived from the functional
integral by assuming the invariance of the fermion measure. As usual
the anomalies can be thought as a non-trivial Jacobian associated
with a transformation of the form (\ref{psitransf}) and therefore in
perturbation theory corresponds to a one-loop effect. For this
reason  we can restrict our attention to one-loop contributions. The
presence of anomalies modifies the first term on the right-hand side
of (\ref{no-anomal}), which acquires a non vanishing value $\delta
W[A]_{anom}$. Below we shall focus on the part $\delta
W[A]_{anom}^{(2)}$ of this functional, which depends quadratically
on $A_{\mu}^b$. To understand the effect of Lorentz violations on
the anomalies one should deal with the 3-point functions and compute
the corresponding triangle graphs (that lead to $\delta
W[A]_{anom}^{(2)}$). These diagrams involve the
complete fermion propagator, which can be obtained by inverting the
operator $i\Gamma^{\mu}D_{\mu} -M$ in (\ref{action}), and the
generalized Dirac matrices $\Gamma^{\mu}$ in the vertices. By
performing this quantum computation in an explicit momentum cutoff
regularization,  one finds that the anomalous part of the WIs is
independent of the Lorentz violating parameters in (\ref{Gansatz})
and (\ref{Mdef}). Moreover, one can explicitly verify that the
anomalous part of $\delta W[A]$ cannot be canceled by adding local
counterterms (which corresponds to a change of the regularization)
even if these counterterms violate  Lorentz symmetry. Therefore, the
anomaly cancellation conditions turn out to be remarkably stable
under the Lorentz violating perturbations that we have considered.

\section{A {\it semiclassical} argument}
The results that we have just summarized come from a detailed
quantum computation performed in Ref. \cite{Salvio:2008ta}. Here we
want to provide a simple argument which shortly illustrates the
origin of the above-mentioned results. Such an argument uses some
quantum results, like the symmetries of $\delta W[A]$, and some
classical aspects, like the structure of the action in
(\ref{action}). We shall therefore refer to it as a semiclassical
argument.

Also in the following we consider the case $$m=0 \quad \mbox{and} \quad H^{\mu
\nu}=0,$$ and so
\be M=a_{\mu}\gamma^{\mu}+ b_{\mu}
\gamma_5\gamma^{\mu}.\label{ab}\ee
Indeed, any term in $M$, which involves an even number of Dirac
matrices, does not contribute to the anomalies and to show this one
can use an argument that leads to the {\it m}-independence of the
anomalies in the Lorentz invariant case \cite{Salvio:2008ta}. We
therefore ignore $m$ and $H^{\mu \nu}$ and refer to
\cite{Salvio:2008ta} for their explicit treatment.

Let us start by considering the anomalous part of $\delta W[A]$ in
the Lorentz invariant case. This may be written \cite{Zumino:1983rz}
as follows:
\be \delta W[A]_{\,anom} = \frac{1}{48\pi^2}  \mbox{Tr}\int d^4x
\,\epsilon^{\mu \nu \lambda \rho} \Omega_L \,\partial_{\mu} \left(2
A_{\nu}^{L} \partial_{\lambda}A_{\rho}^L -i
A_{\nu}^LA_{\lambda}^LA_{\rho}^L\right)- (L\rightarrow R),
\label{LIdeltaW} \ee
where we have defined $$\Omega_{L(R)} \equiv \Omega^b t^b_{L(R)},
\qquad   A_{\mu}^{L(R)} \equiv A_{\mu}^b t^b_{L(R)},$$ and
$\epsilon^{\kappa \nu \lambda \rho}$ is the totally antisymmetric
quantity with $\epsilon^{0123}=1$. The part in (\ref{LIdeltaW}) that
is quadratic in $A^{L(R)}_{\mu}$,
\be \delta W[A]^{(2)}_{\,anom} = \frac{1}{24\pi^2}  \mbox{Tr}\int
d^4x  \,\epsilon^{\mu \nu \lambda \rho} \Omega_L \,\partial_{\mu}
A_{\nu}^{L} \partial_{\lambda}A_{\rho}^L - (L\rightarrow R),
\label{LIdeltaW2} \ee
can be computed by evaluating the 3-point functions
\cite{Bell:1969ts} in a particular regularization, whereas the
remaining terms can be obtained by using the Wess-Zumino consistency
condition \cite{Wess:1971yu}.

We would like to understand why the Lorentz violating deformations
of the theory, corresponding to (\ref{Gansatz}) and (\ref{ab}),
cannot remove the anomalies altogether and to do so it is sufficient
to focus on the quadratic functional in (\ref{LIdeltaW2}). Indeed,
if it were possible to remove the anomalies, in particular there
would be a way to cancel its bilinear part in $\delta W[A]$. Notice
now that (\ref{LIdeltaW2}) is invariant under the following
transformations:
\begin{description}
 \item[1.] Constant shifts of the gauge fields performed independently in the left-handed and right-handed parts,
\item[2.] General coordinate transformations performed independently in the left-handed and right-handed parts.
\end{description}
The first property is generically broken by the cubic terms in (\ref{LIdeltaW}), whereas the second one is
a well-known feature of the complete functional $\delta W[A]_{anom}$.

Meanwhile, the classical action in (\ref{action}) can be expanded as
follows:
\bea S&=& \int d^4x \,i\, \overline{\psi_L(x)}  L^{\mu}_{\,\,\,\nu}
\gamma^{\nu} \left[D_{\mu} + iL^{(-1)\rho}_{\qquad
\mu}\left(a_{\rho}- b_{\rho}\right)\right]\psi_L(x) \nonumber \\&&
+\int d^4x \,i\,  \overline{\psi_R(x)} R^{\mu}_{\,\,\,\nu}
\gamma^{\nu} \left[D_{\mu} +  iR^{(-1)\rho}_{\qquad
\mu}\left(a_{\rho}+
b_{\rho}\right)\right]\psi_R(x),\label{lagrangian}\eea
where $$L^{\mu}_{\,\,\,\nu}\equiv
c^{\mu}_{\,\,\,\nu}-d^{\mu}_{\,\,\,\nu},\quad  R^{\mu}_{\,\,\,\nu}\equiv
c^{\mu}_{\,\,\,\nu}+d^{\mu}_{\,\,\,\nu}$$ and $L^{(-1)\mu}_{\qquad
\nu}$ and $R^{(-1)\mu}_{\qquad \nu}$ are the respective inverse
matrices (which exist in our frame because the breaking of Lorentz invariance has to be small). Also, for later convenience, we have explicitly written
the dependence of the fields on $x$.  The only differences between
(\ref{lagrangian}) and its Lorentz invariant limit are therefore
{\it (i)} two (generically) independent constant shifts of the gauge
fields (due to CPT violating terms in the action)  and {\it (ii)}
two (generically) independent  non singular linear and homogeneous
transformations of $\gamma^{\mu}$ (which can be interpreted as
coordinate transformations).

We can eliminate these differences in the final result for the
anomaly by redefining the gauge fields and the space-time
coordinates as follows. First introduce the new Lie-algebra valued
vector fields
\be A_{\mu}'^{L} \equiv A_{\mu}^L - L^{(-1) \rho}_{\qquad
\mu}\left(a_{\rho}- b_{\rho}\right), \quad A_{\mu}'^{R} \equiv
A_{\mu}^R - R^{(-1) \rho}_{\qquad \mu}\left(a_{\rho}+
b_{\rho}\right).\nonumber \ee
This redefinition serves to hide the CPT-odd parameters $a_{\mu}$
and $b_{\mu}$: now the action can be written in the following way:
\bea S= \int d^4x \,i\, \overline{\psi_L(x)}  L^{\mu}_{\,\,\,\nu}
\gamma^{\nu} D'_{\mu} \psi_L(x) +\int d^4x \,i\,
\overline{\psi_R(x)} R^{\mu}_{\,\,\,\nu} \gamma^{\nu}
D'_{\mu}\psi_R(x),\eea
where $$D'_{\mu}\psi_L (x) \equiv \left(\partial_{\mu} - i
A'^L_{\mu}(x)\right) \psi_L(x)\quad  \mbox{and} \quad D'_{\mu}\psi_R (x) \equiv
\left(\partial_{\mu} - i A'^R_{\mu}(x)\right) \psi_R(x).$$  Then we
consider the additional transformations of the coordinates and the
gauge fields
\bea L^{\mu}_{\,\,\,\nu} \frac{\partial}{\partial x^{\mu}}\equiv
\frac{\partial}{\partial x_{\scriptscriptstyle L}^{\nu}}, \quad
L^{\mu}_{\,\,\,\nu} A'^L_{\mu}(x) \equiv
\tilde{A}^L_{\nu}(x_{{\scriptscriptstyle L}}), \nonumber \\
R^{\mu}_{\,\,\,\nu} \frac{\partial}{\partial x^{\mu}}\equiv
\frac{\partial}{\partial x_{\scriptscriptstyle R}^{\nu}}, \quad
R^{\mu}_{\,\,\,\nu} A'^R_{\mu}(x) \equiv
\tilde{A}^R_{\nu}(x_{{\scriptscriptstyle R}}),\eea
in a way that we can write
\be S= \int d^4x \, i \,\overline{\tilde{\psi}_L(x)} \gamma^{\mu}
\left(\frac{\partial}{\partial x^{\mu}} - i \tilde{A}_{\mu}^L (x)
\right) \tilde{\psi}_L(x) + (L\rightarrow R),\ee
where
\be \tilde{\psi}_L(x) \equiv \sqrt{\mbox{det}L}\, \psi_L (Lx) \quad
\mbox{and}\quad \tilde{\psi}_R(x) \equiv \sqrt{\mbox{det}R}\, \psi_R
(Rx), \label{chiral} \ee
 which is a sort of chiral transformation. We can see that the action assumes a Lorentz invariant form in terms of the new fields $\tilde{\psi}_L$, $\tilde{\psi}_R$, $\tilde{A}^L_{\mu}$ and $\tilde{A}^R_{\mu}$. Therefore, we can derive the anomalies with a standard procedure and,  in a certain regularization, obtain
\be \delta W[A]_{\,anom} = \frac{1}{48\pi^2}  \mbox{Tr}\int d^4x
\,\epsilon^{\mu \nu \lambda \rho} \Omega_L \,\partial_{\mu} \left(2
\tilde{A}_{\nu}^{L} \partial_{\lambda}\tilde{A}_{\rho}^L -i
\tilde{A}_{\nu}^L\tilde{A}_{\lambda}^L\tilde{A}_{\rho}^L\right)-
(L\rightarrow R) \label{tildedeltaW} \ee
(the detailed quantum computation of \cite{Salvio:2008ta} shows that
the possible non invariance of the fermion measure under
(\ref{chiral}) does not lead to corrections). This can be considered
as a generalization of the discussion in \cite{Coleman:1998ti},
where the rotational invariant case has been studied. The fields
$\{\tilde{A}^L_{\mu}, \tilde{A}^R_{\mu}\}$ and $\{A^L_{\mu},
A^R_{\mu}\}$ are related by transformations of the form {\bf 1} and
{\bf 2}, so
\be \delta W[A]^{(2)}_{anom} = \frac{1}{24\pi^2}  \mbox{Tr}\int d^4x
\,\epsilon^{\mu \nu \lambda \rho} \Omega_L \,\partial_{\mu}
A_{\nu}^{L} \partial_{\lambda}A_{\rho}^L - (L\rightarrow R) + ...\,,
\ee
where the dots represent additional bilinear terms coming from the
cubic terms in (\ref{tildedeltaW}), which are not invariant under
constant shifts of the gauge fields. These additional bilinear terms
contain only one derivative and therefore cannot help to cancel
(\ref{LIdeltaW2}), which instead has two derivatives. Not even a
general change of regularization, which does not necessarily assume
Lorentz invariance, can change the fact that $\delta
W[A]_{anom}^{(2)}$ is non-vanishing if it is so in the Lorentz
invariant limit \cite{Salvio:2008ta}.

\section{Conclusions}

In this article we have summarized the discussion about the effect of Lorentz violation on the perturbative non-gravitational anomalies, which has been performed in Ref. \cite{Salvio:2008ta}. Following the SME,  we have assumed locality, translational invariance and power counting renormalizability and focused on a single family of fermions. Although the {\it anomaly functional} $\delta W[A]_{anom}$ can assume a more general form (Lorentz violating counterterms are allowed), the standard anomaly cancellation conditions turn out to be necessary also in the presence of Lorentz violation. Moreover, here we have provided a simple and hopefully illuminating semiclassical argument, which explains the origin of this fact. This relies on two main points. The first one is the analysis of the differences between  the classical structure of the action in the Lorentz invariant and Lorentz violating setups (in the latter case the parameters $L^{\mu}_{\,\,\,\nu}, R^{\mu}_{\,\,\,\nu}, a_{\mu}$ and $b_{\mu}$ are turned on). The second one is the observation of the symmetries of $\delta W[A]^{(2)}_{anom}$, the quadratic part of the anomaly functional. 

An interesting development of this work may be the extension to anomalies that do not correspond to purely internal symmetries, like the gravitational anomalies. Since Lorentz transformations are particular general coordinate transformations, the breaking of Lorentz symmetry should occur spontaneously in the context of gravitational theories, triggered by the vacuum expectation value of tensors \cite{Kostelecky:2003fs}. We therefore expect the generalization to gravitational anomalies to be non-trivial.

\section*{Acknowledgments}
 The author gratefully acknowledges valuable discussions with
Mikhail Shaposhnikov. This work has been supported by the Tomalla
Foundation and by CICYT-FEDER-FPA2008-01430.

\end{document}